\input harvmac
\input epsf.tex
\def\N{{\cal N}}

\noblackbox
%
%
 \def\slash#1{\setbox0=\hbox{$#1$}           
 \dimen0=\wd0                                 
 \setbox1=\hbox{/} \dimen1=\wd1               
 \ifdim\dimen0>\dimen1                        
 \rlap{\hbox to \dimen0{\hfil/\hfil}}      
  #1                                        
 \else                                        
 \rlap{\hbox to \dimen1{\hfil$#1$\hfil}}   
 /                                         
 \fi}                                         %

\def\ev#1{\langle#1\rangle}
\

\lref\IW{
K.~Intriligator and B.~Wecht,
``The exact superconformal R-symmetry maximizes a,''
Nucl.\ Phys.\ B {\bf 667}, 183 (2003)
[arXiv:hep-th/0304128].
}
\lref\IWbar{
K.~Intriligator and B.~Wecht,
``Baryon charges in 4D superconformal field theories and their AdS  duals,''
Commun.\ Math.\ Phys.\  {\bf 245}, 407 (2004)
[arXiv:hep-th/0305046].
}
\lref\BarnesJJ{
  E.~Barnes, K.~Intriligator, B.~Wecht and J.~Wright,
  ``Evidence for the strongest version of the 4d a-theorem, via a-maximization
  along RG flows,''
  Nucl.\ Phys.\ B {\bf 702}, 131 (2004)
  [arXiv:hep-th/0408156].
}

\lref\BGIWii{E. Barnes, E. Gorbatov, K. Intriligator, and J. Wright, ``Current Correlators and AdS/CFT geometry."
}
\lref\JEHO{
  J.~Erdmenger and H.~Osborn,
  ``Conserved currents and the energy-momentum tensor in conformally  invariant
  theories for general dimensions,''
  Nucl.\ Phys.\ B {\bf 483}, 431 (1997)
  [arXiv:hep-th/9605009].
}
\lref\BHac{
  S.~Benvenuti and A.~Hanany,
  ``New results on superconformal quivers,''
  arXiv:hep-th/0411262.
}
\lref\JJN{
I.~Jack, D.~R.~T.~Jones and C.~G.~North,
``$N=1$ supersymmetry and the three loop anomalous dimension for the chiral
superfield,''
Nucl.\ Phys.\ B {\bf 473}, 308 (1996)
[arXiv:hep-ph/9603386];
I.~Jack, D.~R.~T.~Jones and C.~G.~North,
``Scheme dependence and the NSVZ beta-function,''
Nucl.\ Phys.\ B {\bf 486}, 479 (1997)
[arXiv:hep-ph/9609325].
}
\lref\KutasovXU{
  D.~Kutasov and A.~Schwimmer,
  ``Lagrange multipliers and couplings in supersymmetric field theory,''
  Nucl.\ Phys.\ B {\bf 702}, 369 (2004)
  [arXiv:hep-th/0409029].
}
\lref\Anselmi{
  D.~Anselmi,
  ``Central functions and their physical implications,''
  JHEP {\bf 9805}, 005 (1998)
  [arXiv:hep-th/9702056].
}
\lref\ISmir{
  K.~A.~Intriligator and N.~Seiberg,
  ``Mirror symmetry in three dimensional gauge theories,''
  Phys.\ Lett.\ B {\bf 387}, 513 (1996)
  [arXiv:hep-th/9607207].
}
\lref\ISmirr{
  O.~Aharony, A.~Hanany, K.~A.~Intriligator, N.~Seiberg and M.~J.~Strassler,
  ``Aspects of N = 2 supersymmetric gauge theories in three dimensions,''
  Nucl.\ Phys.\ B {\bf 499}, 67 (1997)
  [arXiv:hep-th/9703110].
}
\lref\GW{D.~J.~Gross and F.~Wilczek,
``Asymptotically Free Gauge Theories. 2,''
Phys.\ Rev.\ D {\bf 9}, 980 (1974).
}

\lref\BZ{T.~Banks and A.~Zaks,
``On The Phase Structure Of Vector - Like Gauge Theories With Massless Fermions,''
Nucl.\ Phys.\ B {\bf 196}, 189 (1982).
}

\lref\NSd{
N.~Seiberg,
``Electric - magnetic duality in supersymmetric nonAbelian
gauge theories,''Nucl.\ Phys.\ B {\bf 435}, 129
(1995)[arXiv:hep-th/9411149].
}

\lref\Witten{
  E.~Witten,
  ``Anti-de Sitter space and holography,''
  Adv.\ Theor.\ Math.\ Phys.\  {\bf 2}, 253 (1998)
  [arXiv:hep-th/9802150].
}
\lref\Maldacena{
  J.~M.~Maldacena,
  ``The large N limit of superconformal field theories and supergravity,''
  Adv.\ Theor.\ Math.\ Phys.\  {\bf 2}, 231 (1998)
  [Int.\ J.\ Theor.\ Phys.\  {\bf 38}, 1113 (1999)]
  [arXiv:hep-th/9711200].
}
\lref\PKFL{
  P.~Kraus and F.~Larsen,
  ``Attractors and black rings,''
  arXiv:hep-th/0503219.
}
\lref\GubserVD{
  S.~S.~Gubser,
  ``Einstein manifolds and conformal field theories,''
  Phys.\ Rev.\ D {\bf 59}, 025006 (1999)
  [arXiv:hep-th/9807164].
}
\lref\GKP{
  S.~S.~Gubser, I.~R.~Klebanov and A.~M.~Polyakov,
  ``Gauge theory correlators from non-critical string theory,''
  Phys.\ Lett.\ B {\bf 428}, 105 (1998)
  [arXiv:hep-th/9802109].
}
\lref\GMSW{
  J.~P.~Gauntlett, D.~Martelli, J.~F.~Sparks and D.~Waldram,
  ``A new infinite class of Sasaki-Einstein manifolds,''
  arXiv:hep-th/0403038.
}
\lref\GatesNR{
  S.~J.~Gates, M.~T.~Grisaru, M.~Rocek and W.~Siegel,
  ``Superspace, Or One Thousand And One Lessons In Supersymmetry,''
  Front.\ Phys.\  {\bf 58}, 1 (1983)
  [arXiv:hep-th/0108200].
}
\lref\AnselmiXK{
  D.~Anselmi,
  ``Anomalies, unitarity, and quantum irreversibility,''
  Annals Phys.\  {\bf 276}, 361 (1999)
  [arXiv:hep-th/9903059].
}
\lref\HEK{
  C.~P.~Herzog, Q.~J.~Ejaz and I.~R.~Klebanov,
 ``Cascading RG flows from new Sasaki-Einstein manifolds,''
  JHEP {\bf 0502}, 009 (2005)
  [arXiv:hep-th/0412193].
}
\lref\FMMR{
  D.~Z.~Freedman, S.~D.~Mathur, A.~Matusis and L.~Rastelli,
  ``Correlation functions in the CFT($d$)/AdS($d+1$) correspondence,''
  Nucl.\ Phys.\ B {\bf 546}, 96 (1999)
  [arXiv:hep-th/9804058].
}
\lref\MHKS{
  M.~Henningson and K.~Skenderis,
  ``The holographic Weyl anomaly,''
  JHEP {\bf 9807}, 023 (1998)
  [arXiv:hep-th/9806087].
}
\lref\HOsc{
  H.~Osborn,
  ``N = 1 superconformal symmetry in four-dimensional quantum field theory,''
  Annals Phys.\  {\bf 272}, 243 (1999)
  [arXiv:hep-th/9808041].
}
\lref\WeinbergKK{
  S.~Weinberg,
  ``Charges From Extra Dimensions,''
  Phys.\ Lett.\ B {\bf 125}, 265 (1983).
}
\lref\MS{
  D.~Martelli and J.~Sparks,
  ``Toric geometry, Sasaki-Einstein manifolds and a new infinite class of
  AdS/CFT duals,''
  arXiv:hep-th/0411238.
}

\lref\BBC{
  M.~Bertolini, F.~Bigazzi and A.~L.~Cotrone,
  ``New checks and subtleties for AdS/CFT and a-maximization,''
  JHEP {\bf 0412}, 024 (2004)
  [arXiv:hep-th/0411249].
}
\lref\BHK{
  D.~Berenstein, C.~P.~Herzog and I.~R.~Klebanov,
  ``Baryon spectra and AdS/CFT correspondence,''
  JHEP {\bf 0206}, 047 (2002)
  [arXiv:hep-th/0202150].
}
\lref\HM{
  C.~P.~Herzog and J.~McKernan,
  ``Dibaryon spectroscopy,''
  JHEP {\bf 0308}, 054 (2003)
  [arXiv:hep-th/0305048].
}
\lref\HW{
  C.~P.~Herzog and J.~Walcher,
  ``Dibaryons from exceptional collections,''
  JHEP {\bf 0309}, 060 (2003)
  [arXiv:hep-th/0306298].
}
\lref\DKlm{
D.~Kutasov,
 ``New results on the 'a-theorem' in four dimensional supersymmetric field
theory,''
arXiv:hep-th/0312098.
}

\lref\GSTii{
  M.~Gunaydin, G.~Sierra and P.~K.~Townsend,
  ``Gauging The D = 5 Maxwell-Einstein Supergravity Theories: More On Jordan
  Algebras,''
  Nucl.\ Phys.\ B {\bf 253}, 573 (1985).
}
\lref\Herzog{
  C.~P.~Herzog,
  ``Exceptional collections and del Pezzo gauge theories,''
  JHEP {\bf 0404}, 069 (2004)
  [arXiv:hep-th/0310262].
}
\lref\HKO{
  C.~P.~Herzog, I.~R.~Klebanov and P.~Ouyang,
  ``D-branes on the conifold and N = 1 gauge / gravity dualities,''
  arXiv:hep-th/0205100.
}

\lref\GK{
  S.~S.~Gubser and I.~R.~Klebanov,
  ``Baryons and domain walls in an N = 1 superconformal gauge theory,''
  Phys.\ Rev.\ D {\bf 58}, 125025 (1998)
  [arXiv:hep-th/9808075].
}
\lref\MSY{
  D.~Martelli, J.~Sparks and S.~T.~Yau,
  ``The geometric dual of a-maximisation for toric Sasaki-Einstein manifolds,''
  arXiv:hep-th/0503183.
}
\lref\DuffCC{
  M.~J.~Duff, C.~N.~Pope and N.~P.~Warner,
  ``Cosmological And Coupling Constants In Kaluza-Klein Supergravity,''
  Phys.\ Lett.\ B {\bf 130}, 254 (1983).
}
\lref\GRW{
  M.~Gunaydin, L.~J.~Romans and N.~P.~Warner,
  ``Compact And Noncompact Gauged Supergravity Theories In Five-Dimensions,''
  Nucl.\ Phys.\ B {\bf 272}, 598 (1986).
}
\lref\BF{
  S.~Benvenuti, S.~Franco, A.~Hanany, D.~Martelli and J.~Sparks,
  ``An infinite family of superconformal quiver gauge theories with
  Sasaki-Einstein duals,''
  arXiv:hep-th/0411264.
}
\lref\JEDF{
  J.~Erlich and D.~Z.~Freedman,
  ``Conformal symmetry and the chiral anomaly,''
  Phys.\ Rev.\ D {\bf 55}, 6522 (1997)
  [arXiv:hep-th/9611133].
}
 \lref\AEFJ{D.~Anselmi, J.~Erlich, D.~Z.~Freedman and A.~A.~Johansen,
``Positivity constraints on anomalies in supersymmetric gauge
theories,''
Phys.\ Rev.\ D {\bf 57}, 7570 (1998)
[arXiv:hep-th/9711035].
}
\lref\GST{
  M.~Gunaydin, G.~Sierra and P.~K.~Townsend,
  ``The Geometry Of N=2 Maxwell-Einstein Supergravity And Jordan Algebras,''
  Nucl.\ Phys.\ B {\bf 242}, 244 (1984).
}
\lref\AFGJ{D.~Anselmi, D.~Z.~Freedman, M.~T.~Grisaru and A.~A.~Johansen,
``Nonperturbative formulas for central functions of supersymmetric gauge
theories,''
Nucl.\ Phys.\ B {\bf 526}, 543 (1998)
[arXiv:hep-th/9708042].
}
\lref\deWitEQ{
  B.~de Wit and H.~Nicolai,
  ``N=8 Supergravity With Local SO(8) X SU(8) Invariance,''
  Phys.\ Lett.\ B {\bf 108}, 285 (1982).
}
\lref\HOAP{
  H.~Osborn and A.~C.~Petkou,
  ``Implications of conformal invariance in field theories for general
  dimensions,''
  Annals Phys.\  {\bf 231}, 311 (1994)
  [arXiv:hep-th/9307010].
}
\lref\BJ{M. Baker and K. Johnson, ``Applications of conformal symmetry in quantum electrodynamics,"
Physics 96A (1979) 120.}
\lref\FMMR{
  D.~Z.~Freedman, S.~D.~Mathur, A.~Matusis and L.~Rastelli,
  ``Correlation functions in the CFT($d$)/AdS($d+1$) correspondence,''
  Nucl.\ Phys.\ B {\bf 546}, 96 (1999)
  [arXiv:hep-th/9804058].
}
\lref\Cardy{
  J.~L.~Cardy,
  ``Is There A C Theorem In Four-Dimensions?,''
  Phys.\ Lett.\ B {\bf 215}, 749 (1988).
}
\def\drawbox#1#2{\hrule height#2pt
             \hbox{\vrule width#2pt height#1pt \kern#1pt \vrule
width#2pt}
                   \hrule height#2pt}

\def\Fund#1#2{\vcenter{\vbox{\drawbox{#1}{#2}}}}
\def\Asym#1#2{\vcenter{\vbox{\drawbox{#1}{#2}
                   \kern-#2pt       
                   \drawbox{#1}{#2}}}}

\def\sym{\Fund{6.5}{0.4} \kern-.5pt \Fund{6.5}{0.4}}
\Title{\vbox{\baselineskip12pt\hbox{hep-th/0507137}
\hbox{UCSD-PTH-05-09}}} {\vbox{\centerline{The Exact Superconformal
R-symmetry Minimizes $\tau _{RR}$}}} \centerline{ Edwin Barnes, Elie
Gorbatov, Ken Intriligator, Matt Sudano, and Jason Wright}
\bigskip
\centerline{Department of Physics} \centerline{University of
California, San Diego} \centerline{La Jolla, CA 92093-0354, USA}

\bigskip
\noindent
We present a new, general constraint which, in principle, determines the superconformal
$U(1)_R$ symmetry of 4d $\N =1$ SCFTs, and also 3d $\N =2$ SCFTs.
Among all possibilities, the superconformal $U(1)_R$ is that which minimizes the coefficient, $\tau _{RR}$, of its two-point function.  Equivalently, the superconformal $U(1)_R$ is the unique one
with vanishing two-point function with every non-R flavor symmetry.  For 4d $\N =1$ SCFTs, $\tau _{RR}$ minimization gives an alternative to a-maximization.  $\tau _{RR}$ minimization also applies in 3d, where  no condition
 for determining the superconformal $U(1)_R$ had been previously known.   Unfortunately, this constraint seems
impractical to implement for interacting field theories.  But it can be readily implemented in the AdS geometry
for SCFTs with AdS duals.

\Date{June 2005}

\newsec{Introduction}

Our interest here will be in the coefficients $\tau _{IJ}$ of two-point functions of globally conserved currents $J^\mu _I$ ($I$ labels the various currents) in d-dimensional CFTs:
\eqn\jjev{\ev{J^\mu _I(x)J^\nu _J(y)}={\tau _{IJ}\over (2\pi )^d}(\partial ^2\delta ^{\mu \nu}-\partial ^\mu \partial ^\nu){1\over (x-y)^{2(d-2)}}.}
The general form \jjev\ of the correlator is completely fixed by conformal invariance, with the specific
dynamics of the theory entering only in the coefficients $\tau _{IJ}$.  Unitarity restricts $\tau _{IJ}$ to
be a positive matrix  (positive eigenvalues).   For 4d CFTs, $\tau _{IJ}$ give \refs{\JEHO, \Anselmi}\ the violation of scale invariance, $\ev{T_\mu ^\mu }={1\over 4}\tau _{IJ}(F^I)_{\mu \nu}(F^J)^{\mu \nu}$, when the global currents are coupled to background gauge fields.

We'll here consider field theories with four supercharges: $\N =1$ in 4d, and $\N =2$ in 3d
(one could also consider $\N =(2,2)$ in 2d), and their renormalization group fixed point SCFTs (where there are an additional four superconformal supercharges).  The stress tensor of these theories lives in a supermultiplet
$T_{\alpha \dot \beta}(x, \theta, \overline \theta)$ (in 4d Lorentz spinor notation; for $d<4$ the dot on $\dot \beta$
is unnecessary), which also contains a $U(1)_R$ current -- this is ``the superconformal $U(1)_R$
symmetry".  Supersymmetry relates this current and its divergence to the dilitation current and
its divergence.   The scaling dimension of chiral operators are related to their superconformal
$U(1)_R$ charge by
\eqn\delR{\Delta = {d-1\over 2}R.}
For a chiral superfield, writing $\Delta = \half d -1+\half \gamma$, with $\gamma$ the anomalous dimension, \delR\ yields
\eqn\delRgam{R={d-2\over d-1}+{1\over d-1}\gamma.}

There are often additional non-R flavor currents, whose charges we'll write as $F_i$, with $i$ labeling the flavor symmetries.  In superspace, these currents reside in a different kind of supermultiplet, which we'll write as $J_i(x, \theta, \overline \theta)$.  When there are such additional flavor symmetries, the
superconformal $U(1)_R$ of RG fixed point SCFTs can not be determined by the symmetries alone, as
the R-symmetry can mix with the flavor symmetries.  Some additional dynamical information is
then needed to determine precisely which, among all possible R-symmetries, is the superconformal
one, in the $T_{\alpha \dot \beta}$ supermultiplet.

We will here present a new condition that, in principle, completely determines which is the superconformal $U(1)_R$.   We write the most general possible trial R-symmetry as
\eqn\rgen{R_t=R_0+\sum _i s_i F_i,}
where $R_0$ is any initial R-symmetry, and $F_i$ are the non-R flavor symmetries.  The subscript ``$t$" is for ``trial", with the $s_i$ arbitrary
real parameters.  The superconformal R-symmetry, which we'll write as $R$ without the
subscript, corresponds to
some special values $s_i^*$ of the coefficients in \rgen, that we'd like to determine, $R=R_t|_{s_j=s_j^*}$.

As we'll discuss, the fact that the superconformal R-symmetry and
the non-R flavor symmetries reside in different kinds of supermultiplets, implies that their current-current two-point function necessarily vanishes, $\ev{J_R^\mu (x)J_{F_i}^\nu(y)}=0$, i.e.
\eqn\taufr{\tau _{Ri}=0 \qquad\hbox{for all non-R symmetries $F_i$}.}
This condition uniquely characterizes the superconformal R-symmetry among all
possibilities \rgen.  To see this, use \rgen\ to write \taufr\ as
\eqn\taufrs{0=\tau _{Ri}=\tau _{R_ti}|_{s_j=s_j^*}=\tau _{R_0i}+\sum _j s^*_j \tau _{ij}\quad\hbox{for all i}.}
Here $\tau _{R_0i}$ is the coefficient of the $\ev{J^\mu _{R_0}(x)J^\nu _{F_i}(y)}$ current-current two-point function of the currents for $R_0$ and $F_i$, and $\tau _{ij}$ is the coefficient of the $\ev{J^\mu _{F_i}(x)J^\nu _{F_j}(y)}$ of the current-current two-point function for the non-R flavor symmetries $F_i$ and
$F_j$.  The conditions \taufrs\ is a set of linear equations which uniquely determines the $s_j^*$, if the coefficients $\tau _{R_0i}$ and $\tau _{ij}$ are known.    Unitarity implies that the matrix $\tau _{ij}$ is necessarily positive, with non-zero eigenvalues, so it can be inverted, and the solution of \taufrs\ is
\eqn\sjis{s_j^*=-\sum _i (\tau ^{-1})_{ij}\tau _{R_0i}.}

The conditions \taufrs\ can be
phrased as a minimization principle: {\it the exact superconformal R-symmetry is that which
minimizes the coefficient $\tau _{R_tR_t}$ of its two-point function among all trial possibilities \rgen.}
Using \rgen, the coefficient of the trial R-current $R_t$ two-point function is a quadratic
function of the parameters $s_j$:
\eqn\taurrg{\tau _{R_tR_t}(s)=\tau _{R_0R_0}+2\sum _i s_i \tau _{R_0i}+\sum _{ij}s_is_j \tau _{ij}.}
Our result \taufr\ implies that the exact superconformal R-symmetry extremizes this function,
\eqn\taurrge{{\partial \over \partial s_i}\tau _{R_tR_t}(s)|_{s_j=s_j^*}=2\tau _{Ri}=0.}
The unique solution of \taurrge\ is a global minimum of the function \taurrg\ since
\eqn\taurrgdd{{\partial ^2\over \partial s_i\partial s_j}\tau (s)=2\tau _{ij}>0,}
with the last inequality following from unitarity.

The value of $\tau _{R_tR_t}$ at its unique minimum is the coefficient $\tau _{RR}$ of
the superconformal R-current two-point function.  As is well known, supersymmetry relates
this to the coefficient, ``$c$", of the stress tensor two-point function, $\tau _{RR}\propto c$; as
we'll discuss, the proportionality factor is
\eqn\taurrc{\tau _{RR}={(2\pi )^d\over d(d^2-1)(d-2)}C_T \qquad \hbox{or,  for $d=4$,}\quad \tau _{RR}={16\over 3}c.}

$\tau _{RR}$ minimization immediately implies some expected results.  For non-Abelian
flavor symmetry, \taufr\ is automatically satisfied for all flavor currents with traceless generators,
if the superconformal R-symmetry is taken to commute with these generators.  This shows, as expected, that the
superconformal R-symmetry does not mix with such non-Abelian flavor symmetries.  Similarly, \taufr\ is automatically satisfied by any baryonic flavor
currents which are odd under a charge conjugation symmetry, taking the superconformal
$U(1)_R$ to be even under charge conjugation.  So, as expected, the superconformal
$U(1)_R$ does not mix with baryonic symmetries which are odd under a charge conjugation
symmetry.

As a simple example of $\tau _{RR}$ minimization, consider a single, free, chiral
superfield $\Phi$ in $d$ spacetime dimensions.  The R-symmetry can mix with a non-R $U(1)_F$
flavor current, under which $\Phi$ has charge 1 (the ``Konishi current").  Write the general trial R-charges for the scalar and fermion components as
$R(\phi)=R_t$, $R(\psi)=R_t-1$. As we'll review, the free field two-point function of this R-current is
\eqn\tauffrr{\tau _{R_tR_t}={\Gamma({d\over 2})^22^{d-2}\over (d-1)(d-2)}\left({1\over d-2}R_t^2+(R_t-1)^2\right)}
with the two terms the scalar and fermion contributions.  Taking the derivative w.r.t. $R_t$,
\eqn\tauffrf{\tau _{R_tF}={1\over 2}{d\over dR_t}\tau _{R_tR_t}={\Gamma({d\over 2})^22^{d-2}\over (d-1)(d-2)}\left({R_t\over d-2}+R_t-1\right).}
Requiring $\tau _{RF}=0$ then gives the correct result \delRgam, with anomalous dimension $\gamma =0$, for a free chiral superfield in $d$ spacetime dimensions.

The above considerations all apply independent of space-time dimension; they are equally applicable
for 4d $\N =1$ SCFTs as with 3d $\N =2$ SCFTs.  For 4d $\N =1$ SCFTs, there is already a known method for determining the superconformal R-symmetry:  a-maximization \IW.   It was shown in \IW\ that the $s_i^*$ can be determined by a-maximization, maximizing w.r.t. the $s_i$ in \rgen\ the combination of 't Hooft anomalies
\eqn\amax{a_{trial}(R_t)={3\over 32}(3\Tr R_t^3-\Tr R_t),}
(where we decided here to include the conventional normalization prefactor).  For example, for a free 4d
chiral superfield we locally maximize the function
\eqn\aff{a_{trial}(R_t)={3\over 32}(3(R_t-1)^3-(R_t-1)).}
The local maximum of \aff\ is at $R=2/3$, which indeed coincides with the global minimum of \tauffrr, but it's illustrative to see how the functions themselves differ.

a-maximization in 4d is much more powerful than $\tau _{R_tR_t}$ minimization, because one
can use the power of 't Hooft anomaly matching to exactly compute $a_{trial}(R_t)$ \amax,
whereas the current two-point functions $\tau _{R_0i}$ and $\tau _{ij}$ needed for $\tau _{R_tR_t}$
minimization receive quantum corrections.   Actually, once the exact superconformal R-symmetry is known,
 there is a nice way to evaluate $\tau _{ij}$ in terms of 't Hooft anomalies \AFGJ:
 \eqn\rff{\tau _{ij}=-3\Tr RF_i F_j,}
as we'll review in what follows.  (The result \rff\ generally can not be turned around, and used as a way to determine the superconformal $U(1)_R$, because plugging \rgen\ in \rff\ can not always be inverted to solve for the $s^*$.)

In the context of the AdS/CFT correspondence, the criterion \taufrs\ for determining the
superconformal R-symmetry becomes more useful and tractable, because the AdS duality
gives a weakly coupled dual description of $\tau _{R_0i}$ and $\tau_{ij}$: these quantities
become the coefficients of gauge field kinetic terms in the AdS bulk \FMMR.  As we'll discuss
in a separate paper \BGIWii, these coefficients are computable by reducing SUGRA on the corresponding Sasaki-Einstein space.  We'll show in \BGIWii\ that the conditions \taufrs\ are in fact equivalent to the ``geometric dual of
a-maximization" of Martelli, Sparks, and Yau \MSY.

There is no known analog of a-maximization for 3d $\N =1$ SCFTs, and in 3d there is no useful analog of 't Hooft anomalies and matching (aside from a $Z_2$ parity anomaly matching \ISmirr).  $\tau _{R_tR_t}$ minimization gives an alternative to a-maximization in 4d, which applies equally well to 3d $\N =2$ SCFTs.

a-maximization in 4d ties the problem of finding the superconformal $U(1)_R$ together with
Cardy's conjecture \Cardy, that the conformal anomaly $a$ counts the degrees of freedom of a quantum field theory, with $a_{UV}>a_{IR}$ and $a_{CFT}>0$.  The result that $a$ is maximized
over its possibilities implies that relevant deformations decrease $a$ \IW, in agreement with
Cardy's conjecture.  Unfortunately, we have not gained any new insights here into general RG inequalities {}from our $\tau _{RR}$ minimization result.  Indeed, $\tau _{RR}$ is related to the
conformal anomaly $c$ in 4d, which is known to not have any general behavior, neither
generally increasing nor generally decreasing, in RG flows to the IR.  And there is no analogous
argument  to that of \IW, to conclude that $\tau _{RR}$ generally increases in RG flows in the IR, from the fact that $\tau _{RR}$ is minimized among all possibilities: the quantum corrections to $\tau _{RR}$, coming from the relevant interactions, can generally have either sign.  (The difference is that the argument of \IW\ was based on 't Hooft anomalies, which do not get any quantum  corrections for conserved currents).

Our $\tau _{RR}$ minimization result applies for SCFTs at their RG fixed point.  It would be
interesting to extend $\tau _{RR}$ minimization to study RG flows away from the RG fixed
point.  Perhaps this can be done by using Lagrange multipliers, as in \DKlm, to impose the constraint that one minimize only over currents that are conserved by the relevant interactions.

\newsec{Current two point functions; free fields and normalization conventions}

Two point functions of currents and stress tensors for free bosons and
fermions in d-spacetime dimensions were worked out, e.g. in \HOAP.  To compare with \HOAP, rewrite \jjev\ as
\eqn\jjevv{\ev{J^\mu _I(x)J^\nu _J(y)}=\tau _{IJ}{2(d-1)(d-2)\over (2\pi )^d}{I_{\mu \nu}(x-y)\over (x-y)^{2(d-1)}},}
with $I_{\mu \nu}(x)\equiv \delta _{\mu \nu}-2x_\mu x_\nu (x^2)^{-1}$.
The normalization conventions of \HOAP\ is
\eqn\HOconv{\ev{J_\mu (x)J_\nu (0)}={C_V\over x^{2(d-1)}}I_{\mu \nu}(x), \qquad
\ev{T_{\mu \nu}(x)T_{\rho \sigma}(0)}={C_T\over x^{2d}}I_{\mu \nu, \sigma \rho}(x),}
with  $I_{\mu \nu, \sigma \rho}(x)=\half (I_{\mu \sigma}(x)I_{\nu\rho}(x)+I_{\mu \rho}(x)I_{\nu \sigma}(x))-d^{-1}\delta _{\mu \nu}\delta _{\sigma \rho}$.  Thus $C_V=2\tau (d-1)(d-2)/(2\pi )^d$.
With these  normalizations, the coefficients \HOconv\ for a single complex scalar are
\eqn\scvct{C_V={2\over d-2}{1\over S_d^2}, \qquad C_T={2d\over d-1}{1\over S_d^2},}
where $S_d\equiv 2\pi ^{\half d}/\Gamma (\half d)$ and the current was normalized to give
$\phi$ and $\phi ^*$ charges $\pm 1$.  The coefficients for a free fermion having the same number
of components as a 4d complex chiral fermion  (half the components of a Dirac
fermion) the coefficients are
\eqn\fcvct{C_V=2{1\over S_d^2}, \qquad C_T=d{1\over S_d^2}}
(we don't have the factors of $2^{d/2}$ of \HOAP, because we're here considering a fermion
with the same number of components as the dimensional reduction of a 4d chiral fermion for all $d$).

More generally,  let current $J_I(x)$ give charges $q_{I,b}$ to the complex bosons and
charges $q_{I,f}$ to the chiral fermions.  Using \scvct\ and \fcvct,  we have
\eqn\tauijff{ \tau _{IJ}^{\hbox{free field}}={\Gamma ({d\over 2})^22^{d-2}\over (d-1)(d-2)}\left({1\over d-2}\sum _{\hbox{bosons}\ b} q_{I,b}q_{J,b}+\sum_{\hbox{fermions}\ f}
q_{I,f}q_{J,f}\right).}
In particular, for a $U(1)_R$ symmetry, this gives \tauffrr. For $d=4$, $\Gamma (d/2)^22^{d-2}/(d-1)(d-2)=2/3$, so e.g. a 4d $U(1)_F$ non-R symmetry which assigns charge $q$ to a single chiral
superfield has
 $\tau _{FF}^{free field}=q^2$.

\newsec{Supersymmetric field theories}

Supersymmetry relates the superconformal R-symmetry to the stress tensor: both reside in
the supercurrent supermultiplet
\eqn\tmult{T_{\alpha \dot \alpha}(x, \theta , \overline \theta )\sim J_{R,\alpha \dot \alpha}(x)+S_{\alpha \dot \alpha \beta}(x)\theta ^\beta + \overline S_{\alpha \dot \alpha \dot \beta}(x)\overline \theta ^{\dot \beta}+T_{\alpha\dot \alpha \beta \dot \beta}(x) \theta ^\beta \overline \theta{\dot \beta}+\dots,}
whose first component is the superconforal $U(1)_R$ current and whose $\theta\overline \theta$ component is
the stress energy tensor (we're omitting numerical coefficients here).   Our notation is for the 4d case; similar results hold for 3d $\N =2$ theories, with $\overline \theta ^{\dot \alpha}$ replaced with a second flavor of $\theta ^\alpha$.   For superconformal theories, the stress tensor is traceless, and the superconformal R-current is conserved.
As discussed in \HOsc, the supercurrent two-point function is then of a completely determined form, with the only dependence on the theory contained in a single overall coefficient $C$:
\eqn\ttvev{\ev{T_{\alpha \dot \alpha}(z_1)T_{\beta \dot \beta}(z_2)}=C {(x_{1\overline 2})_{\alpha \dot \beta}(x_{2\overline 1})_{\beta \dot \alpha}\over (x_{\overline 21}^2x_{\overline 1 2}^2)^{d/2}};}
see \HOsc\ for an explanation of the superspace notation in \ttvev.

Expanding out \ttvev\ in superspace, the LHS includes both the R-current two-point function and
the stress-tensor two-point function.  So  \ttvev\ shows that the coefficient $C\propto \tau _{RR}$, and also $C\propto C_T$, and so it follows that $\tau _{RR}\propto C_T$.    We could determine the
precise coefficients in these relations by being careful with the coefficients in \tmult\ and
in expanding both sides of \ttvev; instead we will fix these universal proportionality factors by
considering the particular example of a free chiral superfield.   Using \scvct\ and \fcvct\
to get $C_T$, and comparing with the free-field value of $\tau _{RR}$ computed from \tauijff,
gives the general proportionality factor that we quoted in \taurrc; e.g. for $d=3$ it's
$\tau _{RR}=\pi ^3C_T/3$.  In 4d, $C_T\propto c$, one of the conformal anomaly coefficients,
and the proportionality can again be fixed by considering the case of a free 4d $\N=1$ chiral superfield, for which $c=1/24$ and \tauijff\ gives $\tau _{RR}= 2/9$
(or a free 4d $\N =1$ vector superfield,
for which $c=1/8$ and  \tauijff\ gives $\tau _{RR}=2/3$); this gives the relation quoted in \taurrc.

The non-R global flavor currents $J_i^\mu(x)$ are the $\theta ^\alpha \overline \theta ^{\dot \alpha}$ components of superfields $J_{i}(x, \theta, \overline \theta )$, whose
first component is a scalar.  We can write their two-point functions in superspace \HOsc,
with the coefficients given by that of the flavor current correlators,  $\tau _{ij}$:
\eqn\jjss{\ev{J_{i}(z_1)J_j(z_2)}= {\tau _{ij}\over (2\pi )^d}{1\over (x_{\overline 21}^2x_{\overline 12}^2)^{(d-2)/2}}.}

In general $d$ dimensional CFTs, two-point functions of primary operators vanish unless the operators
have conjugate Lorentz spin and the same operator dimension.  Noting that the first component
of the supermultiplet \tmult\ has dimension $\Delta (T_{\alpha \dot \beta})=d-1$, and the first component
of the current $J_i$ has dimension $\Delta (J_i)=d-2$ (since the $\theta ^\alpha \overline \theta ^{\dot \alpha}$ component is the current, with dimension $d-1$), the two-point function of the first components of these two different supermultiplets  must vanish.  Because there is no non-trivial nilpotent invariant for
two-point functions \HOsc , this implies that two-point function of the entire supermultiplets must
vanish:
\eqn\tj{\ev{T_{\alpha \dot \alpha}(z_1)J_i(z_2)}=0.}
I.e. the two-point function of any operator in the $T_{\alpha \dot \alpha}$ supermultiplet and any operator
in the $J_i$ supermultiplet vanishes; in particular, this implies that the two-point function of the
superconformal $U(1)_R$ current and all non-R flavor currents necessarily vanish, $\tau _{RF_i}=0$.
We thus have the general result \taufr, and this same argument applies  equally for $d=4$ $\N =1$ as
well as lower dimensional SCFTs with the same number of supersymmetries.

\subsec{4d $\N =1$ SCFTs: relating current correlators to 't Hooft anomalies}

The superspace version of an anomaly in the dilatation current is
\eqn\sanom{\overline{ \grad{}}^{\dot \alpha}T_{\alpha \dot \alpha}=\grad{}_\alpha L_T,}
with $L_T$ the trace anomaly, which is the variation of the effective action with respect to the
chiral compensator chiral superfield \GatesNR.

On a curved spacetime, there is the conformal anomaly
\eqn\ca{\ev{T^\mu _\mu}={1\over 120}{1\over (4\pi )^2}\left (c(\hbox{Weyl})^2-{a\over 4}(\hbox{Euler})\right),}
(there can also be an $a' \partial ^2R$ term, whose coefficient $a'$ is ambiguous, which was
discussed in detail in \AnselmiXK). The coefficient ``$c$" is that of the stress tensor two-point function in flat space, whereas the coefficient ``$a$" can be related to a stress tensor 3-point function in flat space.  The superspace version of this anomaly, including also background gauge fields coupled to the superconformal R-current,  is as in \sanom, with $L_T=(c{\cal W}^2-a\Xi _c)/24\pi ^2$ \AFGJ.  Taking components of this superspace anomaly equation relates the conformal anomaly coefficients $a$ and $c$ to the 't Hooft anomalies of the
superconformal $U(1)_R$ symmetry \AFGJ:
\eqn\acthooft{a={3\over 32}(3\Tr R^3-\Tr R)\qquad c={1\over 32}(9\Tr R^3-5\Tr R).}

An alternate derivation \HOsc\ of these relations follows from the fact that, in flat space,
the 3-point function $\ev{T_{\alpha \dot \alpha}(z_1)T_{\beta \dot \beta}(z_2)T_{\gamma \dot \gamma}(z_3)}$ is of a form that's completely determined by the symmetries and Ward identities, up to two overall normalization coefficients, with
one linear combination of these coefficients proportional to the coefficient \ttvev\ of the $T_{\alpha \beta}$ two-point function.  In components,
this relates the stress tensor three-point functions, and hence $a$ and $c$, and to the R-current 3-point functions, and hence the $\Tr U(1)_R$ and $\Tr U(1)_R^3$ 't Hooft anomalies, to these
two coefficients.  It follows that $a$ and $c$ can be expressed as linear combinations of
$\Tr U(1)_R$ and $\Tr U(1)_R^3$, and the coefficients in \acthooft\ can easily be determined
by considering the special cases of free chiral and vector superfields.

Combining and \acthooft, we have
\eqn\susyreln{\tau _{RR}={3\over 2}\Tr R^3 -{5\over 6}\Tr R.}

It was also argued in \AFGJ\ that the two-point functions $\tau _{ij}$ of non-R flavor currents
are related to 't Hooft anomalies, as
\eqn\RFF{\tau _{ij}=-3\Tr RF_i F_j.}
Again, this can be argued for either by turning on background fields, or by considering
correlation functions in flat space.  In the former method, one uses the fact that coupling
background field strengths to the non-R currents leads to $\Delta L_T=C_{ij}W_{\alpha i}W^\alpha _j$, in \sanom, for some coefficients $C_{ij}$.  In components, \sanom\ then gives
$\delta \ev{T_\mu ^\mu}\sim C_{ij}F_{\mu \nu, i}F^{\mu \nu}_j$ and $\delta \ev{\partial _\mu J_R^\mu}\sim C_{ij}F_{\mu \nu i}\widetilde F^{\mu \nu}_ j$.  The former gives $C_{ij}\sim \tau _{ij}$
and the latter gives $C_{ij}\sim \Tr RF_iF_j$, so  $\tau _{ij}\propto \Tr RF_iF_j$.  The
coefficient in \RFF\ is again easily determined by considering the special case of free field theory.

The alternate derivation would be to consider the flat space
3-point function of the stress tensor and two flavor currents,  $\ev{T_{\alpha \dot \alpha}(z_1)J_i(z_2)J_j(z_3)}$.  It was shown in \JEHO\ that such 3-point functions are completely determined
by the symmetries and Ward identities, up to two overall coefficients, and that one linear combination of these coefficients is proportional to the current-current two point functions, and
hence $\tau _{ij}$.  In our supersymmetric context, that same linear combination should be related by supersymmetry to $\ev{\partial _\mu J_R^\mu (x_1)J_{F_i}^\rho (x_2)J_{F_j}^\sigma (x_3)}$, and
hence to the $\Tr RF_iF_j$ 't Hooft anomaly.

The a-maximization \IW\ constraint on the superconformal R-symmetry follows from the fact
that supersymmetry relates the $\Tr R^2F_i$ and $\Tr F_i$ 't Hooft anomalies:
\eqn\amaxi{9\Tr R^2F_i-\Tr F_i=0,}
which again can be argued for either by considering again an anomaly with background fields, or by considering current correlation functions in flat space \IW.
In the former method, one considers
the anomaly of the flavor current coming from a curved background metric and background gauge
field coupled to the superconformal R-current,  $\overline{\grad{}} ^2J\propto {\cal W}^2$.
With the latter method, one uses the result of  \HOsc\ that the flat space 3-point
function $\ev{T_{\alpha \dot \alpha}(z_1)T_{\beta \dot \beta}(z_2)J_i(z_3)}$ is completely determined
by the symmetries and superconformal Ward identities, up to a single overall normalization
constant.

We note that supersymmetry does not relate $\tau _{Ri}$ to the 't Hooft anomaly $\Tr R^2F_i$.
Naively, one might have expected some such relation, in analogy with the above arguments,
for example by trying to use \sanom\ to relate a term $\delta \ev{T^\mu _\mu}\sim \tau _{Ri}F_{R,\mu \nu}F_i^{\mu \nu}$ to a term $\delta \ev{\partial _\mu J_R^\mu }\sim (\Tr R^2F_i) F_{R, \mu \nu}\widetilde F_i ^{\mu \nu}$,
when background fields are coupled to both $U(1)_R$ and $U(1)_{F_i}$ currents.  But there is actually no  way to write such combined contributions of the $U(1)_R$ and $U(1)_{F_i}$ background fields to \sanom, because the former resides  in the spin 3/2 chiral super field strength ${\cal W}_{\alpha\beta \gamma}$, and the latter resides in the spin 1/2 chiral super field strength $W_{\alpha i}$, and there is no way to combine the two of them into the spin zero chiral object $L_T$.   Likewise, in flat space, a relation between $\tau _{Ri}$ and $\Tr R^2F_i$ would occur
if the 3-point function $\ev{T_{\alpha \dot \alpha}(z_1)T_{\beta \dot \beta}(z_2)J_i(z_3)}$, which
includes a term proportional to $\Tr R^2F_i$, were related to the two-point function
$\ev{T_{\beta \dot \beta}(z_2)J_i(z_3)}$, which is proportional to $\tau _{Ri}$ (and, as we have
argued above, vanishes).  It sometimes happens that 3-point functions with a stress tensor
are simply proportional to the 2-point function without the stress tensor, e.g. this is the case
when the other two operators are chiral and anti-chiral primary \HOsc.  But the
the  $\ev{T_{\alpha \dot \alpha}(z_1)T_{\beta \dot \beta}(z_2)J_i(z_3)}$ 3-point function in \HOsc\ is not related to the $\ev{T_{\beta \dot \beta}(z_2)J_i(z_3)}$ two-point function.  Indeed, the free field example discussed in the introduction illustrates that $\Tr R^2F_i$ and $\tau_{Ri}$ are not related by supersymmetry,  as $\Tr R^2F_i\neq 0$ for this example but, as always, $\tau _{Ri}=0$.

\subsec{Using $\tau_{Ri}=0$ to determine the superconformal R-symmetry}

As discussed in the introduction, using \rgen, we have for a general trial R-symmetry
\eqn\tautrial{\tau _{R_ti}=\tau _{R_0i}+\sum _j s_j \tau _{ij}.}
Imposing $\tau _{R_ii}=0$ gives a set of linear equations, which determines the particular values
$s_j^*$ of the parameters for which the trial R-symmetry is the superconformal R-symmetry.  As discussed in the introduction, this can equivalently be expressed as ``the exact superconformal
R-symmetry minimizes its two-point function coefficient $\tau _{R_tR_t}(s)$, which is given by \taurrg, and which we can re-write  using $\tau _{Ri}=0$ for the superconformal R-symmetry as
\eqn\taurtrti{\tau _{R_tR_t}(s)=\tau_{RR}+\sum _{ij}(s_i-s_i^*)(s_j-s_j^*)\tau _{ij},}
making it manifest that $\tau _{R_tR_t}$ has a unique global minimum, when the $s_j$ are
set to the particular value $s_j^*$.  At $s_j=s_j^*$, the general R-symmetry $R_t$ in \rgen\ becomes
the superconformal R-symmetry, in the supermultiplet stress tensor $T_{\alpha \dot \alpha}$.

The function $\tau _{R_tR_t}(s)$ to minimize and the function $a_{trial}(s)$ to locally maximize
in 4d are different.  Let us compare the values of them and their derivatives at the extremal point
$s_i=s_i^*$.  For \tautrial, we have:
\eqn\tautrid{\eqalign{\tau _{R_tR_t}|_{s^*}&=\tau _{RR}={16\over 3}c={3\over 2}\Tr R^3-{5\over 6}\Tr R, \cr
{\partial\over \partial s_i}\tau _{R_tR_t}|_{s^*}&=0, \cr {\partial ^2\over \partial s_i\partial s_j}\tau _{R_tR_t}&=2\tau _{ij},}}
whereas for ${16\over 3}a_{trial}(R_t)\equiv \half (3\Tr R_t^3-\Tr R_t)$ we have:
\eqn\atrid{\eqalign{{16\over 3}a_{trial}(R_t)|_{s^*}&={16\over 3}a={3\over 2}\Tr R^3-{1\over 2}\Tr R, \cr
{\partial\over \partial s_i}{16\over 3}a_{trial}(R_t)|_{s^*}&={9\over 2}\Tr R^2F_i-{1\over 2}\Tr F_i=0, \cr {\partial ^2\over \partial s_i\partial s_j}{16\over 3}a_{trial}(R_t)|_{s^*}&=9\Tr RF_iF_j=-3\tau _{ij}.}}
The derivatives of both functions of $s$ vanish at the same values $s^*$.  The values of the two functions in \tautrid\ and \atrid\ differ, except for SCFTs with $a=c$, i.e. $\Tr R=0$, as is the case
for SCFTs with AdS duals \foot{Quite generally, quiver gauge theories with only bi-fundamental matter
 have $\Tr R=0$, and hence $a=c$  \refs{\IWbar , \BHac }. }
 The second derivatives of the functions in \tautrid\ and \atrid\  are
proportional, though with opposite sign, reflecting the fact that the exact superconformal R-symmetry
minimizes $\tau _{R_tR_t}$ and maximizes $a_{trial}(R_t)$.

For the sake of comparison, let's also consider the function ${16\over 3}c_{trial}(R_t)\equiv {3\over 2}R_t^3-{5\over 6}R_t$; the value of this function and its first two derivatives at $R_t=R$, i.e. $s_i=s_i^*$, are
\eqn\ctrid{\eqalign{{16\over 3}c_{trial}(R_t)|_{s^*}&={16\over 3}c={3\over 2}\Tr R^3-{5\over 6}\Tr R, \cr
{\partial\over \partial s_i}{16\over 3}c_{trial}(R_t)|_{s^*}&={9\over 2}\Tr R^2F_i-{5\over 6}\Tr F_i=-{1\over 3}\Tr F_i, \cr {\partial ^2\over \partial s_i\partial s_j}{16\over 3}c_{trial}(R_t)|_{s^*}&=9\Tr RF_iF_j=-3\tau _{ij}.}}
The value of $\tau _{R_tR_t}$ and $c_{trial}(R_t)$ coincide at $R_t=R$.  The value of their first derivatives differ for any flavor symmetries with $\Tr F_i\neq 0$.  General SCFTs can have flavor
symmetries with $\Tr F_i=0$, but SCFTs with AdS duals always have $\Tr F_i=0$, and $\Tr F_i=0$ for general superconformal quivers with only bifundamental matter \refs{\IWbar, \BHac}.   The second
derivatives in \ctrid\ differ from those of \tautrid\ by a factor of $-3/2$, coinciding with those of \atrid.

As a further comparison of $a$-maximization in 4d with $\tau _{RR}$ minimization, let's
consider the equations for the case where the superconformal $U(1)_R$ can mix with
a single non-R flavor symmetry, $R_t=R_0+sF$.  $a$-maximization gives the value $s^*$
for the superconformal $U(1)_R$ as a solution of the quadratic equation
\eqn\quads{s^2\Tr F^3+2s\Tr R_0F^2+\Tr R_0^2F-{1\over 9}\Tr F=0.}
$\tau _{RR}$ minimization gives $s^*$ as \sjis
\eqn\lins{s^*=-\tau _{R_0F}/\tau _{FF}.}
If $\Tr F^3$ is non-zero, $s^*$ can also be obtained from \rff, which
here gives
\eqn\linsx{s^*=-\left[\Tr R_0F^2+{1\over 3}\tau _{FF}\right]/\Tr F^3.}
For any given choice of $R_0$ and $F$, the value of $s^*$ obtained in these three different
ways must agree.  It would be nice to have a direct proof of the relations that this implies.
E.g. comparing  \linsx\ with \lins\ gives the identity $\tau _{R_0F}\Tr F^3=\tau _{FF}\left({1\over 3}\tau _{FF}+\Tr R_0F^2\right)$ which, evidently,  must hold for any choice of the R-symmetry $R_0$ (taking $R_0$ to equal the superconformal $U(1)_R$, both sides vanish).

\newsec{SQCD Example}

4d $\N =1$ SCQD, with gauge group $SU(N_c)$ and $N_f$ fundamental and anti-fundmantal flavors, $Q$ and $\widetilde Q$,  has been argued to flow to a SCFT in the IR for the flavor range ${3\over 2}N_c<N_c<3N_c$ \NSd.  Taking the superconformal $U(1)_R$ to be the anomaly free R-symmetry,
the superconformal R-charges are $R(Q)=R(\widetilde Q)=1-(N_c/N_f)$.  Let's also consider the baryonic $U(1)_B$ symmetry, with $B(Q)=-B(\widetilde Q)=1/N_c$.  Using the 't Hooft anomaly relations,
\eqn\taurrsqcdex{\tau _{RR}={3\over 2}\Tr R^3-{5\over 6}\Tr R={3\over 2}\left[N_c^2-1-2{N_c^4\over N_f^2}\right]+{5\over 6}\left[N_c^2+1\right],}
\eqn\taubbsqcdex{\tau _{BB}=-3\Tr RBB=6.}
For $N_f\approx 3N_c$, where the RG fixed point is at weak coupling as in \refs{\GW, \BZ}, these
expressions reduce to the free field values.

There is a unique, anomaly free $U(1)_R$ symmetry that commutes with charge
conjugation and the $SU(N_f)$ global symmetries.  Our $\tau _{R_tR_t}$
minimization condition immediately leads to the same conclusion.  $\tau _{R_tR_t}$
is minimized by having
$\tau _{RB}=0$ and $\tau _{RF_i}=0$ for the $U(1)_B$ and $SU(N_f)$ global symmetries.
Taking the $U(1)_R$ to be even under charge conjugation ensures that $\tau _{RB}=0$, because
the $U(1)_B$ current is odd, so charge conjugation symmetry gives $\tau _{RB}=-\tau _{RB}$.
Likewise $\tau _{RF_i}=0$ for the $SU(N_f)$ flavor currents, simply by the tracelessness of
the generators, if $U(1)_R$ is taken to commute with $SU(N_f)$.

\newsec{Perturbative analysis}

Consider a general 4d $\N =1$ SCFT with gauge group $G$ and matter chiral superfields $Q_f$
in representations $r_f$ (of dimension $|r_f|$) of $G$, with no superpotential, $W=0$.  If the theory is just barely
asymptotically free, there can be a RG fixed point at weak gauge coupling, where perturbative
results can be valid.  We will verify that the leading order pertubative
expression for the anomalous dimension for fields,
\eqn\adi{\gamma _f(g)=-{g^2\over 4\pi ^2}C(r_f)+{\cal O}(g^4),\qquad\hbox{i.e}\qquad R_f={2\over 3}-{g^2\over 12\pi ^2}C(r_f)+O(g^4).}
agrees with $\tau _{RR}$ minimization.
As standard, we define  group theory factors as
\eqn\grpthy{\Tr _{r_f}(T^AT^B)=T(r_f)\delta ^{AB}, \qquad \sum _{A=1}^{|G|}T_{r_f}^AT^A_{r_f}=C(r_f){\bf 1}_{|r_f|\times |r_f|}, \quad \hbox{so}\quad C(r_f)={|G|T(r_f)\over |r_f|}.}
The RG fixed point value $g_*$ of the coupling is determined by the constraint that the R-symmetry be anomaly free, $T(G)+\sum _f T(r_f)(R_f-1)=0$.

For the free UV theory, we minimize $\tau _{RR}$ over all possible R charges $R_f$
of the matter chiral superfields, which are unconstrained for $g=0$.  As we discussed in
the introduction, this gives the free-field
term $R_f^{(0)}=2/3$.
For $g\neq 0$, we write
$R_f=R_f^{(0)}+R_f^{(1)}+\dots$, with $R_f^{(1)}$ the $O(g^2)$ term that we'd like to find
via $\tau _{RR}$ minimization.  For $g\neq 0$, $\tau _{RR}$ should be minimized subject to
the constraint that the symmetries be anomaly free, i.e. we impose $\tau _{Ri}=0$ over all
anomaly free $U(1)_R$ and $U(1)_{F_i}$ symmetries, with R charges $R_f$,  and flavor $F_i$ charges $q_i(r_f)$ constrained to satisfy
\eqn\anomf{T(G)+\sum _f T(r_f)(R_f-1)=0, \qquad \hbox{and}\qquad \sum _f T(r_f)q_i(r_f)=0.}

The $U(1)_R$ current assigns charges $R_f$ to the squark and $R_f-1$ to the quarks components
of $Q_f$.   The $U(1)_{F_i}$ non-R current assigns charges $q_i(r_f)$ to both the quark
and squark components of $Q_f$.  To compute $\tau _{RF_i}$, we consider the diagrams
for the two point function $\ev{J_R^\mu (x_1)J_{F_i}^\nu (x_2)}$.  Because we take the currents
to be conserved, they have vanishing anomalous dimension, so we anticipate that the various
diagrams sum such that all apparent divergences cancel, and we're left with only  finite
contributions to $\tau _{RF_i}$.  The $O(g^2)$ contributions can be written as
\eqn\taufrpii{\tau _{Ri}^{(1)}=\sum _f q_i(r_f)\left[({1\over 3}R_f^{(1)}+{2\over 3}R_f^{(1)})|r_f|+R_f^{(0)}(A_f^{(1)}+C_f^{(1)})+(R_f^{(0)}-1)(B_f^{(1)}+C_f^{(1)})\right].}
The first two terms come from the leading diagrams, without interactions, exactly as in
the free-field result \tauffrf, but weighted by
the $O(g^2)$ R-charges $R_f^{(1)}$.  The first term is from
 connecting the currents at $x_1$ and $x_2$, with squark $\phi _f$
propagators, and the second from connecting them with quark $\psi _f$ propagators.
The remaining contributions in \taufrpii\ are $O(g^2)$ because they involve
$O(g^2)$ interaction diagrams, and the R-charge weighting is thus taken as
$R^{(0)}=2/3$. Here $A_f^{(1)}$ is the contribution of all $O(g^2)$ 1PI diagrams connecting squark
$\phi _f$, at $x_1$, to squark $\phi _f$ at $x_2$.  $B_f^{(1)}$ is similarly the contribution
{}from all $O(g^2)$ diagrams connecting quark $\psi _f$ at $x_1$ to quark $\psi _f$ at $x_2$.  $C_f^{(1)}$ is the contributions of diagrams connecting squark $\phi _f$ at $x_1$ to quark $\psi _f$ at $x_2$  (or vice-versa).    We note that the
 group theory factors in all of these diagrams with
$O(g^2)$ interactions is the same: $\Tr _{r_f} \sum _{A=1}^{|G|}T^A_{r_f}T^A_{r_f}=|r_f|C(r_f)=|G|T(r_f)$,
i.e. $A_f^{(1)}=|G|T(r_f)A^{(1)}$, $B_f^{(1)}=|G|T(r_f)B^{(1)}$, and $C_f^{(1)}=|G|T(r_f)C^{(1)}$,
where $A^{(1)}$, $B^{(1)}$, and $C^{(1)}$ are independent of the gauge group and representation, e.g. they could be computed in $U(1)$ SQED.

Using the second constraint in \anomf, $\sum _f T(r_f)q_i(r_f)=0$, it immediately follows, without even having to compute $A^{(1)}$, $B^{(1)}$, and $C^{(1)}$, that their contributions to $\tau _{Ri}^{(1)}$ in \taufrpii\ all vanish, for all anomaly free flavor symmetries $F_i$.  The only contributions remaining in
\taufrpii\ are the $R_f^{(1)}$ ones,  $\tau _{Ri}^{(1)}=\sum _f q_i(r_f)R^{(1)}_f|r_f|$.  Our $\tau _{RR}$ minimization result implies that this must vanish, for any $q_i(r_f)$ satisfying the
anomaly free constraint in \anomf.  This implies that $R_f^{(1)}=\alpha C(r_f)$ for some
constant $\alpha$ that's independent of the rep. $r_f$.

We have thus used $\tau _{R_tR_t}$ minimization to re-derive the group theory dependence
of the $O(g^2)$ term in the anomalous dimension \adi.  The coefficient is also fixed to agree
with \adi, at the fixed point $g_*$, by using the condition in \anomf\ that
the R-symmetry be anomaly free to solve for $\alpha$ (which is appropriately
small when the matter content is such that the theory is barely asymptotically free).  This reproduces the $O(g^2)$ contribution to the R-charges in \adi\ at the RG fixed point.

In principle, one could extend this analysis, and use $\tau _{RR}$ minimization to compute
the anomalous dimensions to all orders.  Using a-maximization \IW\ (assuming that the
RG fixed point has no accidental symmetries), the general result can be written as \DKlm
 \eqn\adrgf{R_f=  {2\over 3}(1+{1\over 2}\gamma _f(g_*))=1-{1\over 3}\sqrt{1+{\lambda_*T(r_f)\over |r_f|}}=1-{1\over 3}\sqrt{1+{\lambda _*C(r_f)\over |G|}},}
where $\lambda _*$ is a Lagrange multiplier \DKlm, which is determined by the constraint that the R-symmetry be anomaly free, $T(G)+\sum _f T(r_f)(R_f-1)=0$. The result \adrgf\ was successfully
compared \refs{\BarnesJJ, \KutasovXU}\ with the results for the anomalous dimensions to 3-loops of \JJN.   But, because current two-point functions get quantum corrections, $\tau _{RR}$ minimization
does not seem to be a very efficient way to compute anomalous dimensions.  Indeed,
the higher order quantum corrections to $\tau _{Ri}$ include diagrams where
the currents at $x_1$ and $x_2$ are connected by renormalized propagators, with all quantum corrections from the interactions, and computing such $\tau _{Ri}$ contributions is already tantamount to directly computing the anomalous dimensions $\gamma _f(g)$.

\centerline{\bf Acknowledgments}

KI thanks the ICTP Trieste and CERN for hospitality during
the final stage of writing up this work, and the groups and visitors there for discussions.
 This work was supported by DOE-FG03-97ER40546.

\listrefs\end
\appendix{A}{Computing the $O(g^2)$ contributions to $\tau _{Ri}$.}

As we discussed above, we can recover the $O(g^2)$ anomalous dimension from
$\tau _{Ri}=0$ simply from the group theory dependence of the diagrams, and the
constraint that $U(1)_R$ and $U(1)_{F_i}$ be anomaly free, without having to actually
compute the $O(g^2)$ perturbative corrections to $\tau _{Ri}$.  Nevertheless, let us here
sketch the computation of these corrections.

One way to do the computation is to work in position
space, using the additional freedom to do conformal inversions $x_\mu = x'_\mu /x'{}^2$, in analogy with the current three-point function analysis of \JEDF.  The Jacobian is $\partial x_\mu /\partial x'_\nu = x^2(\delta _{\mu \nu}-2x_\mu x_\nu /x^2)\equiv x^2I_{\mu \nu}(x)$.  The scalar and spinor propagators
transform as
\eqn\spfpc{\eqalign{\Delta (x-y)&={1\over 4\pi ^2}{1\over (x-y)^2}={1\over 4\pi ^2}{x'{}^2y'{}^2\over (x'-y')^2}\cr
S(x-y)=-\slash{\partial} \Delta (x-y)&={1\over 2\pi ^2}{\slash{x}-\slash{y}\over (x-y)^4}=-{1\over 2\pi ^2}x'{}^2y'{}^2\slash{x}'{(\slash{x}'-\slash{y}')\over (x'-y')^2}\slash{y}'.}}
For the gauge field \refs{\BJ\JEDF}\ propagator
\eqn\gfprop{\Delta _{\mu \nu}={1\over 4\pi ^2}\left[{\delta _{\mu \nu}\over (x-y)^2}-\half \Gamma {I_{\mu \nu}(x-y)\over (x-y)^2}\right],}
with $\Gamma =0$ in Feynman gauge and $\Gamma =1$ in Landau gauge.

Let us consider the diagram with the R-current at points $x_1$ and $x_2$, connected by fermion
propagators, with a gluon exchange between the fermion propagators:
\eqn\fdex{g^2\int d^4ud^4v S(x_1-u)S(u-x_2)S(x_1-v)S(v-x_2)\Delta _{\mu \nu}(u-v)\delta ^{\mu \nu}.}

\listrefs

\end